\documentclass[allclo]{FBSart}
\usepackage{amsfonts}
\usepackage{amssymb}
\usepackage{graphicx}
\newcommand{\nabsq}{\overrightarrow{\nabla}^{2}\!}
\newcommand{\beq}{\begin{equation}}
\newcommand{\eeq}{\end{equation}}
\newcommand{\bea}{\begin{eqnarray}}
\newcommand{\eea}{\end{eqnarray}}

\title{Universality in Few-Body Systems}
\author{L.~Platter
\thanks{\textit{Former address:} 
Department of Physics and Astronomy, Ohio University,
Athens, OH 45701, USA}}
\institute{
Department of Physics, Ohio State University,
Columbus, OH 43210, USA}

\runningauthor{L.~Platter}
\runningtitle{Universality in Few-Body Physics}
\begin{document}

\maketitle
\begin{abstract}
Low-energy universality in atomic few-body systems as a
result of a large two-body scattering length has gained
a lot of attention recently. Here, I discuss recent progress
in describing the three-body recombination of cold atoms
in terms of a finite set of universal scaling functions
and review results for the recombination length of
$^{133}$Cs atoms obtained with these functions.
Furthermore, I will consider the inclusion of effective range
corrections and the relevance for further calculations in atomic
and nuclear physics.
\end{abstract}

\section{Introduction}
\label{sec:intro}
Non-relativistic three-body systems of identical bosons with a large
two-body scattering length display a rich set of universal features.
In particular, if $a=\pm\infty$ the system has infinitely many
3-body bound states ({\it trimers}) with an accumulation point at the scattering
threshold. In the zero-range limit the binding energies of these
trimers are given by
\beq
\label{eq:efimov}
B_3^{(n)}=(e^{-2\pi/s_0})^{n-n_*}\kappa_*^2/m~,
\eeq
where $\kappa_*$ is the binding wavenumber of the branch 
of Efimov states labeled by $n_*$ and $s_0 \approx 1.00624$
in the case of identical bosons.

Efimov who found these results pointed out that these universal
features are also valid in the case of large but finite scattering
length as long as it remains large compared to the range $R$ of
the underlying interaction \cite{Efimov70}.
The implications of these discoveries are manifold, in particular
as they are independent of the details of the interaction and
therefore apply to systems of nucleons but also to cold atoms
which have an unnaturally large scattering length. These findings
have become even more important over the last years since Efimov's
results have been rephrased in terms of an effective field theory (EFT).
It turns out that this EFT is the appropriate low-energy theory
for non-relativistic particles with short-range interactions:
It allows model-independent calculation of observables, the
inclusion of external currents is straightforward and corrections
due to the finite range of the interaction can be included
systematically. This framework should therefore be ideally suited
to compute electroweak few-nucleon observables which are for
example of interest in nuclear astrophysics
($p\,d\,\rightarrow \gamma\,^3$He). 

The leading order (LO) implications of this EFT also known as
Efimov physics are interesting by itself. In nuclear
physics these consequences can be seen in correlations
between different observables. If for example the binding
energy of the triton is computed with different
nucleon-nucleon potentials which reproduce the same
two-body observables but give different values for the
neutron-deuteron scattering length, a plot of these
atom-dimer scattering lengths versus the corresponding results
for the triton binding energy will give an approximately
linear correlation which is known as the Phillips line. 

The first experimental evidence for Efimov physics 
has recently been presented by the Innsbruck group \cite{Grimm06}.
They carried out experiments with ultracold 
$^{133}$Cs atoms, using a magnetic field to control
their scattering length. They observed a resonant enhancement
in the three-body recombination rate at $a \approx -850 \ a_0$
that can be attributed to an Efimov trimer close to the three-atom threshold.
The scattering length dependent loss rate at 10~nK can be fit 
well by the universal formula for zero temperature derived
in \cite{Braaten:2003yc}.

In the following, I will discuss several applications of
the EFT for short-range interactions relevant to the
themes mentioned above. I will
briefly review the theoretical basis on which this EFT
is built in section \ref{sec:eft}.
Then I will discuss recent work on universal
functions which allow to parameterize the three-body
recombination of identical bosons. In
section \ref{sec:range}, I will consider how to account
for finite range corrections within this framework
and will give results for observables which have been
calculated up to next-to-next-to-leading order (NNLO)
in the EFT expansion. I will end with a short conclusion.

\section{An EFT for Short-Range Interactions} 
\label{sec:eft}
The EFT for short-range interactions is formulated in terms of the
minimal set of degrees of freedom, {\it i.e.} heavy boson or fermion fields
only and is valid if the underlying potential is short-ranged and the
involved momentum is smaller than the inverse range of the potential.
The most general Lagrangian describing non-relativistic bosons interacting
through contact interactions only is given by

\bea
  {\cal L}  &=&
       \psi^\dagger \biggl[i\partial_t + \frac{\nabsq}{2M}\biggr]
                 \psi - \frac{C_0}{2}(\psi^\dagger \psi)^2
            - \frac{D_0}{6}(\psi^\dagger\psi)^3
                                           +  \ldots
\,,\label{lag}
\eea
where the ellipses denote interactions with more derivatives
and/or more fields.
With the corresponding power counting this EFT is an expansion in $R/a$,
where $R$ denotes the range of the underlying potential and $a$ the
two-body scattering length. It is a particular feature of this theory
that when applied to the three-body sector, a three-body force
is needed at leading order for the correct renormalization
of observables \cite{Bedaque:1998kg}.
Thus, one three-body datum is needed to fix the three-body
low-energy constant in order to make predictions for the
remaining observables.
At leading order, observables will display universality
which means that they only depend on the scattering length and
an additional three-body parameter. 
Instead of using a three-body force one can use alternatively
a subtraction formalism which trades the three-body force in the
kernel of the three-body Faddeev equation for an explicit
three-body observable in the inhomogeneity of the integral equation
\cite{Afnan:2003bs,Hammer:2000nf}. 
It was shown recently that within this framework no
energy-dependent three-body force needs to be included up to
next-to-next-to-leading order (NNLO) \cite{Platter:2006ev}. 
\section{Universal Functions}
\label{sec:universal}
Universality in the three-body system implies that observables
can be written as a function of the scattering length $a$ and
a three-body parameter. This statement can be made more precise
and in Efimov's radial law the S-matrix elements for three-body scattering
are written as a combination of universal scaling functions and complex
phase factors which only contain the three-body parameter \cite{Braaten:2004rn}
%%%%%%%%%%%%%%%%%%%%%%%%%%%%%%%%%%% 
\begin{figure}
\centerline{\includegraphics*[width=9.8cm,angle=0,clip=true]{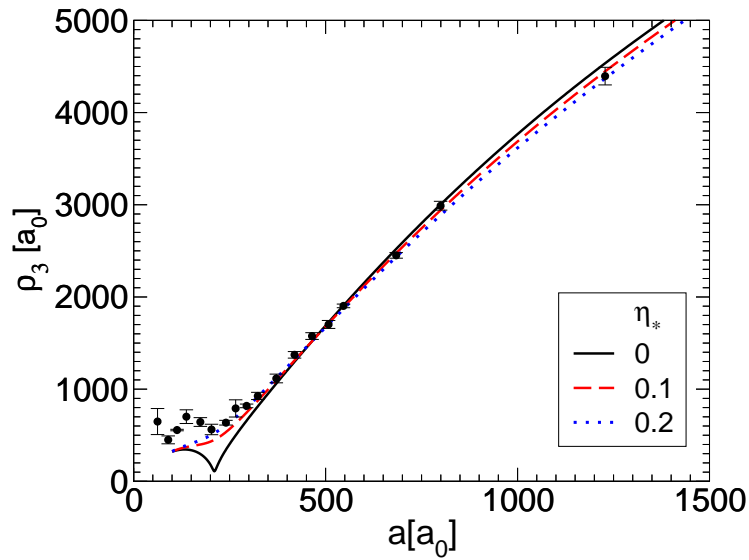}}
\caption{\label{fig:rho3}The 3-body recombination length $\rho_3$ 
for $^{133}$Cs atoms as a function of $a$ for $T=200$ nK.
The data points are from Ref.~\cite{Grimm06}.
The curves are the universal prediction for three values 
of $\eta_*$: 0 (solid line), 0.1 (dashed line), and 0.2 (dotted line).}
\end{figure}
%%%%%%%%%%%%%%%%%%%%%%%%%%%%%%%%%%%
\bea
\label{eq:radial_law}
\nonumber
S_{AD,AD}&=&s_{22}(x)+\frac{s_{21}(x)s_{12}(x)\,e^{2i s_0\ln
  (a/a_{*0})}}{1-s_{11}(x)\,e^{2i s_0\ln
  (a/a_{*0})}}~,\\
\nonumber
S_{AD,AAA}&=&s_{23}(x)+\frac{s_{21}(x)s_{13}(x)\,e^{2i s_0\ln
  (a/a_{*0})}}{1-s_{11}(x)\,
e^{2i  s_0\ln (a/a_{*0})}}~,\\
S_{AAA,AAA}&=&s_{33}(x)+\frac{s_{31}(x)s_{13}(x)\,e^{2i s_0\ln
  (a/a_{*0})}}{1-s_{11}(x)\,
e^{2i  s_0\ln (a/a_{*0})}}~,
\eea
where $AAA$ ($AD$) denotes a final/initial states of three atoms
(an atom and a dimer) and the functions $s_{ij}$ depend on the scaling
variable $x=\sqrt{E m a^2}$. The three-body parameter $a_{*0}$ is determined
by the scattering length at which the three-body recombination rate
displays a minimum.

In \cite{Braaten:2006qx}, it was pointed out that
calculations using conventional potentials displaying a large two-body
scattering length can be used to extract these universal functions.
However, only one such calculation for the recombination $^4$He atoms
was available at that time and various simplifying assumptions were made
to reduce the number of functions to one. 
In Refs.~\cite{Shepard:2007gj} the recombination rate of $^4$He atoms
was calculated with several potentials and the simplifying assumptions
were successively relaxed so that the
3-body recombination rate was determined by two \cite{Shepard:2007gj}
and then three \cite{Platter:2007sn} independent scaling functions, 
respectively.

The EFT for short-range interactions allows a precise determination
of the universal functions given in Eq.(\ref{eq:radial_law}).
The amplitude for elastic atom-dimer scattering can easily be
calculated for a wide range of energies and three-body parameters
and this in turn allows to fit all universal functions relevant
for three-body recombination precisely \cite{universal2007}.

The knowledge of these functions allows to calculate the
scattering length dependent three-body recombination length $\rho$ 
of $^{133}$Cs atoms as measured by the Innsbruck group.
The excellent agreement between the universal prediction for the
recombination length $\rho_3$ and the experimental results
can be seen in Fig.\ref{fig:rho3}.
The figure contains the universal predictions for three different
values of the parameter $\eta_*$ which quantifies the impact
of additional recombination channels into deep dimers. These
deep bound states cannot be described within the EFT for
short-range interactions and their effects are not universal.
\section{Range Corrections}
\label{sec:range}
%%%%%%%%%%%%%%%%%%%%%%%%%%%%
\begin{figure}
\centerline{
\includegraphics*[width=10cm,angle=0,clip=true]{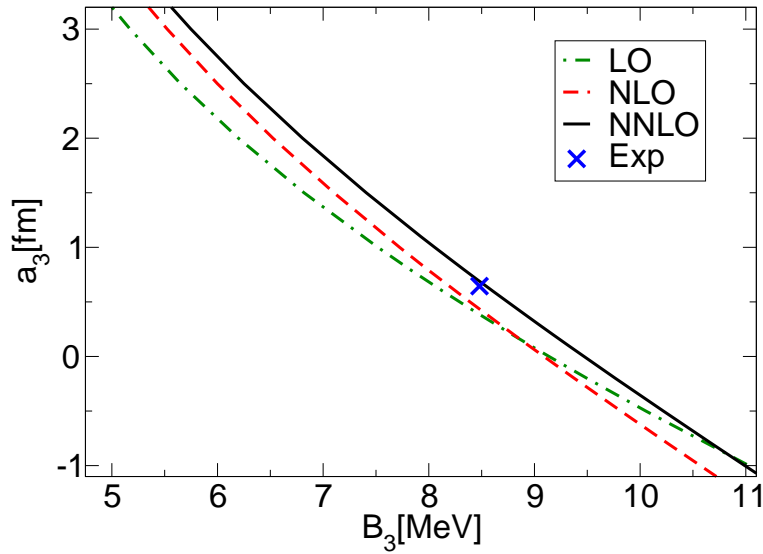}}
\caption{\label{fig:phillips} The correlation between the neutron-deuteron
scattering length and the triton binding energy. The curves represent
the results at LO (dot-dashed), NLO (dashed line) and NNLO (solid line). 
The cross gives the experimental result.}
\end{figure}
%%%%%%%%%%%%%%%%%%%%%%%%%%%%
In experiments with cold atoms which exploit Feshbach resonances,
the scattering length can often only be varied within a certain range.
Depending on the maximal value of the scattering length or on
the energy scale of the observable of interest it might
therefore be of interest to account for finite range effects in
the calculation. Furthermore, in nuclear physics the ratio
$R/a\sim 1/3$ is fixed and therefore range corrections have to be
taken into account for a precise determination of low-energy
three-body observables. The EFT for short-range interactions allows
to account for range effects systematically. 
The impact of range corrections at next-to-leading order (NLO) was considered
in \cite{Hammer:2001gh} and observables in the three-nuclon
sector were computed up to NNLO in \cite{Bedaque:2002yg}.
In the latter work explicit three-body counterterms were used
and at NNLO an additional energy-dependent counterterm
was introduced for the renormalization of observables at this order.
In \cite{Platter:2006ev} it was shown that observables can be computed
up to NNLO with no additional three-body parameter if a subtraction
is employed to account for the three-body input parameter at
leading order. This subtraction formalism has been used to compute
the binding energies and scattering phaseshifts of the $^4$He
atomic three-body system and the three-nucleon system up to NNLO.
The results are in excellent agreement with experiment and previous
calculations using conventional atom-atom/nucleon-nucleon interactions.
For example, in \cite{Platter:2006ad} the triton binding energy
is calculated to be $B_t^{(\hbox{\tiny LO})}=8.08$~MeV,
$B_t^{(\hbox{\tiny NLO})}=8.19$~MeV
at LO and NLO, respectively and $B_t^{(\hbox{\tiny NNLO})}=8.54$~MeV at NNLO if the 
neutron-deuteron scattering length $a_{nd}=0.65$~fm \cite{Huffman:2005jx}
is used as three-body input parameter. This is in very good agreement with the
experimental value of the triton binding energy $B_t^{\rm exp}=8.48$~MeV.
The correlation between the neutron-deuteron scattering length and the
triton binding energy (the Phillips line) is displayed in Fig.~\ref{fig:phillips}. 
\section{Summary}
\label{sec:summary}
The EFT for short-range interactions is a valuable tool to compute
low-energy observables of nuclear and atomic few-body systems.
Recently, this EFT has been used to extract a set of universal functions
(defined in Efimov's radial law) which allow to parameterize
the three-body recombination rate in terms of the previously
defined scaling parameter $x$ and the three-body parameter $a_{*0}$.
These functions were then used to compute the recombination rate for
$^{133}$Cs atoms at positive scattering length at non-zero temperature.
The results are in excellent agreement with experimental data obtained
by the Innsbruck group. 

In the three-nucleon system the inclusion of higher order corrections is
necessary to achieve this level of agreement between theory and experiment.
I have discussed some recent results for observables in the three-nucleon
for which range corrections have been taken into account \cite{Platter:2006ad}.
A small number of interesting calculations in which the three-body
nucleon system is coupled to external currents has already been
performed\cite{Platter:2005sj,griesshammer2006}.
This gives reason to hope that further
observables relevant to nuclear astrophysics will be computed
with this EFT in the near future.
%Finally, I want to point out that this EFT
%also seems to be the ideal tool to analyze the Ab-problem in 
%nucleon-deuteron scattering as it is a model-independent
%framework which is -- through its low-energy constants --
%intimately connected to low-energy scattering data.

\begin{acknowledge}
This work was supported in part by the Department of Energy 
under grant DE-FG02-93ER40756, by the National Science Foundation
under Grant No.~PHY--0653312, by an Ohio University postdoctoral
fellowship and the UNEDF SciDAC Collaboration under DOE Grant
DE-FC02-07ER41457.
\end{acknowledge}

\end{document}